\documentclass[twocolumn,aps,prl,showpacs,superscriptaddress,groupedaddress]{revtex4}
\usepackage{textcomp}
\usepackage{amsmath}
\usepackage{amssymb}
\usepackage{graphicx}
\usepackage{color}

\begin{document}

\title{Ultra-low-noise room-temperature quantum memory for polarization qubits.}
\author{Mehdi Namazi, Connor Kupchak, Bertus Jordaan, Reihaneh Shahrokhshahi and Eden Figueroa}
\affiliation{Department of Physics and Astronomy, Stony Brook University, New York 11794-3800, USA}

\begin{abstract}
Here we show an ultra-low noise regime of operation in a simple quantum memory in warm $^{87}$Rb atomic vapor. By modelling the quantum dynamics of four-level room temperature atoms, we achieve fidelities $>$90\% for single-photon level polarization qubits, clearly surpassing any classical strategy exploiting the non-unitary memory efficiency. This is the first time such important threshold has been crossed with a room temperature device. Additionally we also show novel experimental techniques capable of producing fidelities close to unity. Our results demonstrate the potential of simple, resource-moderate experimental room temperature quantum devices.
\end{abstract}
\maketitle

Robust and operational room temperature quantum devices are a fundamental cornerstone towards building quantum networks composed of a large number of light-matter interfaces \cite{Ritter2015,Borregaard2016}. Such quantum networks will be the basis of the creation of quantum repeater networks \cite{DLCZ} and measurement device independent quantum cryptography links \cite{Abruzzo2014,Panayi2014}. Given the recent success in the creation of elementary playgrounds in which single photons interact with atoms in controlled low temperature environments \cite{Lvovsky2009,Bussieres2013,Northup2014,Reiserer2014,Heshami2016}, the next technological frontier is the design of interfaces where such phenomena can be performed without extra-cooling \cite{Lee2011,Bader2014,Yao2012,Krauter2013,Jensen2016}. The big challenge for such room temperature operation is to defeat the inherent strong atomic motion, decoherence and a considerable amount of background photons present \cite{Eisaman2005, Reim2011, Hosseini2011_2, Sprague2014,Manz2007,Phillips2011,Bashkansky2012,Vurgaftman2013}. A pertinent metric of these effects is the SBR, defined as $\eta/q$, where $\eta$ is the retrieved fraction of a single excitation stored in a quantum memory and $q$ the average number of concurrently emitted photons due to background processes.

\noindent \emph{Quantum memory setup and storage parameters optimization.} Our experimental setup includes four aspects of utmost relevance in order to allow for high SBR and quantum memory fidelity at the single-photon level:\\
\noindent a) Dual rail operation. We store pulses containing on average one qubit in warm $^{87}$Rb vapor using electromagnetically induced transparency (EIT). Two independent control beams coherently prepare two volumes within a single $^{87}$Rb vapor cell at $60^{\circ}\,$C, containing Kr buffer gas, thus serving as the storage medium for each mode of a polarization qubit. We employed two external-cavity diode lasers phase-locked at 6.835 GHz. The probe field frequency is stabilized to the $5S_{1/2} F = 1$ $\rightarrow$ $5P_{1/2} F' = 1$ transition at a wavelength of 795 nm (detuning $\Delta$) while the control field interacts with the $5S_{1/2} F = 2$ $\rightarrow$ $5P_{1/2} F' = 1$ transition.\\
\noindent b) Control field suppression. Polarization elements supply 42 dB of control field attenuation (80\% probe transmission) while two temperature-controlled etalon resonators (linewidths of 40 and 24 MHz) provide additional 102 dB. The total probe field transmission is 4.5\% for all polarization inputs, exhibiting an effective, control/probe suppression ratio of 130 dB.\\
\noindent c) Background/efficiency compromise. The storage efficiency and the number of background photons posses different dependence on the control field power. Optimal qubit storage fidelities are obtained for non-maximal storage efficiency. A combination of these three techniques have been used in our previous investigation to obtain fidelities $>$75\% with storage efficiencies $\sim$ 5\% \cite{Kupchak2015}.\\
\noindent d) Probe temporal duration. Best impedance matching between the field and the EIT storage medium is achieved by temporal shaping of the probe field pulses. In our previous work we have experimentally characterized the optimal temporal bandwith of the probe photons to be $\sim$ 500 ns, using feed-forward cascaded storage \cite{Namazi2015}.\\
\begin{figure*}
\centerline{\includegraphics[width=2.0\columnwidth]{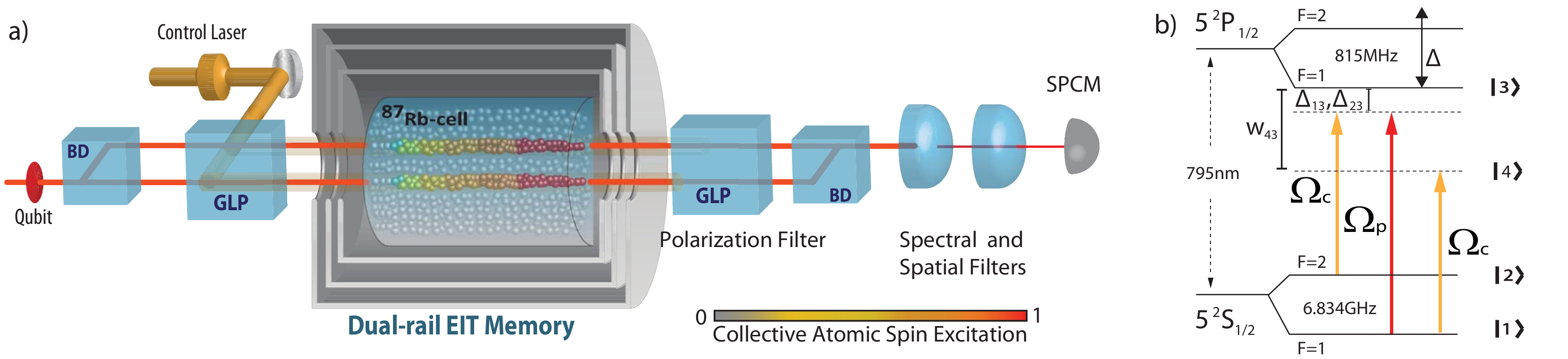}}
\caption{(a)Dual-rail quantum memory setup. Probe: red beam paths; control: yellow beam paths; BD: Polarization Beam Displacer; GLP: Glan Laser Polariser; SPCM: Single Photon Counter Module. The color-code bar depicts the strength of the collective atomic excitation. (b) Rubidium D1 line four-level scheme describing the transitions used in the description of the efficiency and background response. $|1\rangle$ and $|2\rangle$ (ground states), $|3\rangle$ (excited state), $|4\rangle$ (off-resonant virtual state) and $\Delta$ (one-photon laser detuning).}
\end{figure*}
\noindent \emph{Full quantum model of room temperature operation.} Surpassing any classical strategy exploiting non-unitary memory efficiencies requires increasing the SBR substantially. To do so we have developed a model of the quantum dynamics of the room temperature quantum memory. We start by considering atoms exhibiting a four-level energy level scheme interacting with two laser fields, $\Omega_{p}$ (probe) and $\Omega_{c}$ (control), with one-photon detunings $\Delta_{13}$ and $\Delta_{23}$ respectively (see Fig. 1b). We include the off-resonant interaction of the control field with a virtual state $|4\rangle$. The phenomenological Hamiltonian describing the atom-field coupling in a rotating frame is:
\begin{equation}
\begin{split}
\hat{H}=(-\Delta_{13}+\Delta)\hat{\sigma}_{11}-(\Delta_{13}-\Delta_{23})\hat{\sigma}_{22} \\ -\Omega_{p}E_{p}\hat{\sigma}_{31}-\Omega_{c}E_{c}\hat{\sigma}_{32}-\frac{\alpha}{\omega_{43}+\Delta}\Omega_{c}E_{c}\hat{\sigma}_{41}- \\ \frac{\alpha}{\omega_{43}+\Delta}\Omega_{c}E_{c} \hat{\sigma}_{42}-(\Delta_{13}-\omega_{43})\hat{\sigma}_{44} +h.c,
\nonumber
\end{split}
\end{equation}
where $\Delta$ is the laser detuning, $\alpha$ is the coupling strength to the virtual state, $\hat{\sigma}_{ij}=|i\rangle\langle j|,i,j=1,2,3,4$ are the atomic raising and lowering operators for $i\neq j$, and the atomic energy-level population operators for $i=j$ and ${E}_{p}(z,t)$ and ${E}_{c}(z,t)$ are the normalized electric field amplitudes of the probe and control fields. We use the master equation:
\begin{equation}
\begin{split}
\dot{\hat{\rho}}=-i[\hat{H},\hat{\rho}]+\sum_{m=1,2}\Gamma_{3m}(2\hat{\sigma}_{m3}\rho\hat{\sigma}_{3m}-\hat{\sigma}_{33}\hat{\rho}-\hat{\rho}\hat{\sigma}_{33})\\
+\sum_{m=1,2}\Gamma_{4m}(2\hat{\sigma}_{m4}\rho\hat{\sigma}_{4m}-\hat{\sigma}_{44}\hat{\rho}-\hat{\rho}\hat{\sigma}_{44})\\
+\Gamma_{12}(2\hat{\sigma}_{21}\rho\hat{\sigma}_{12}-\hat{\sigma}_{11}\hat{\rho}-\hat{\rho}\hat{\sigma}_{11})
\nonumber
\end{split}
\end{equation}
together with the Maxwell-Bloch equation, $\partial_{z}{E}_{p}(z,t)=i\frac{\Omega_{p}N}{c}\langle\hat{\sigma}_{31}(z,t)\rangle$, to calculate the expected retrieved pulse shape $E_{OUT}(t)$ and the storage efficiency bandwidth response $\eta(\Delta)$. Here $L$ is the atomic sample length, $\Gamma$'s being the decay rates of the excited levels, $c$ is the speed of light in vacuum and $N$ the number of atoms.\\
\begin{figure}[h]
\includegraphics[width=\linewidth]{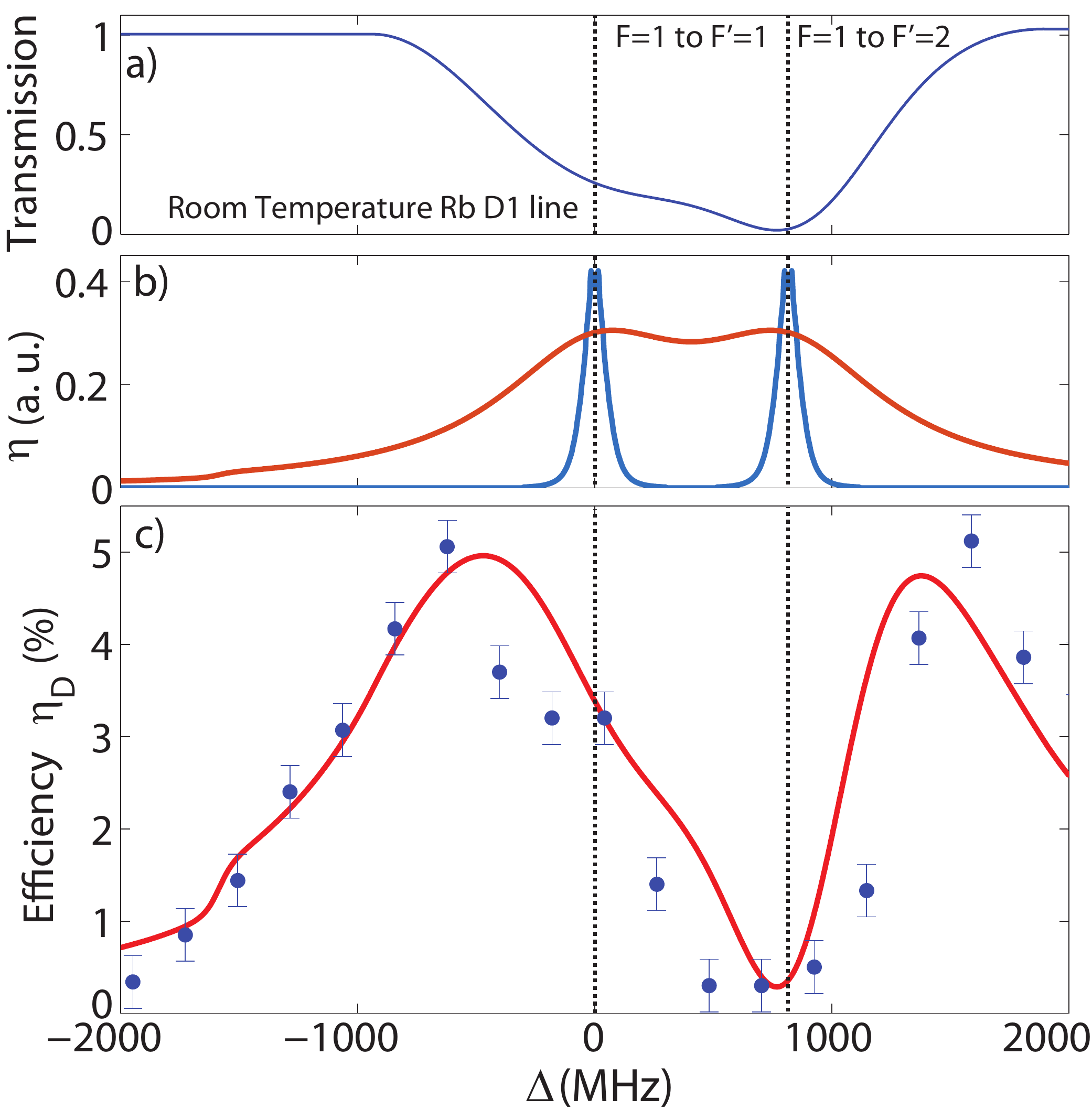}
\caption{(a) Measured transmission profile $T_{RT}(\Delta)$. (b) Cold atom storage bandwidths $\eta_{1}(\Delta)$ and  $\eta_{2}(\Delta)$ for the two excited states of the rubidium D1 line manifold (the blue line is a master equation prediction of the storage bandwidth) and room temperature storage bandwidth $\eta_{RT}(\Delta)$ (the solid red line is the result of the convolution with a velocity distribution). (c) Overall efficiency response $\propto (\eta_{RT}(\Delta))(T_{RT}(\Delta))$ (solid red line) and storage experiments over a 4 GHz scan region with a central frequency at the $F=1$ to $F'=1$ D1 line rubidium transition (blue dots). The error bars are statistical.}
\end{figure}
The room temperature response is calculated by convolving two storage efficiency bandwidths $\eta_{1}(\Delta)$ and $\eta_{2}(\Delta)$ (corresponding to two excited states in the rubidium D1 line manifold, blue line in Fig. 2b) with a distribution $A(\Delta)=A({2 \pi v}/{\lambda})=\frac{\sqrt{\ln 2}}{W_d \sqrt{\pi}}\frac{1}{1+(2\Delta)^2/W_d^2}$. We have set $W_d$ to 960MHz to include also pressure broadening effects (obtained from a fit on the measured transmission profile (Fig. 2a)). Defining $\Delta=\Delta_{j} = \Delta_{0} + j\Delta_{step}$ we calculate the response as $\eta(\Delta_{j})=\sum_{i=-i_{max}}^{i_{max}}A(\Delta_{i})\eta(\Delta_{ j + i})$. The resultant broadened storage bandwidth $\eta_{RT}(\Delta)$ is presented in Fig. 2b (solid red line). We also account for the varying optical depth at different $\Delta$ by multiplying $\eta_{RT}(\Delta)$ by the measured transmission profile $T_{RT}(\Delta)$ (see Fig. 2a). The resultant is the room temperature efficiency bandwidth (see Fig. 2c red line).\\
We perform storage experiments for $1/\sqrt{2}(|H\rangle~+~|V\rangle)$ qubits with a storage time of 700~ns over a $\Delta$ region of 4 GHz. Figure~2c compares these results to our model. The most striking observation is that the maximum storage efficiency is not achieved on atomic resonance, but at detunings beyond the Doppler width. The maximum efficiencies are at $\Delta$~=~500~MHz (red detuned) and $\Delta$=~1.3~GHz (blue detuned).\\
\noindent \emph{Single-photon level background reduction.} Having found non-trivial regions of optimal operation, we now simulate the quantum dynamics of the atomic system when no probe field is present. The contribution of the Stokes field in the memory background is calculated using an extra term to ${E}_{p}(z,t)$ relative to $\langle\hat{\sigma}_{42}(z,t)\rangle$. The numerical values used are $\Gamma_{3m}=3MHz$, $\Gamma_{4m}=1GHz$ and the decoherence rate between ground states 0.1kHz.
\begin{figure}[h!]
\includegraphics[width=\linewidth]{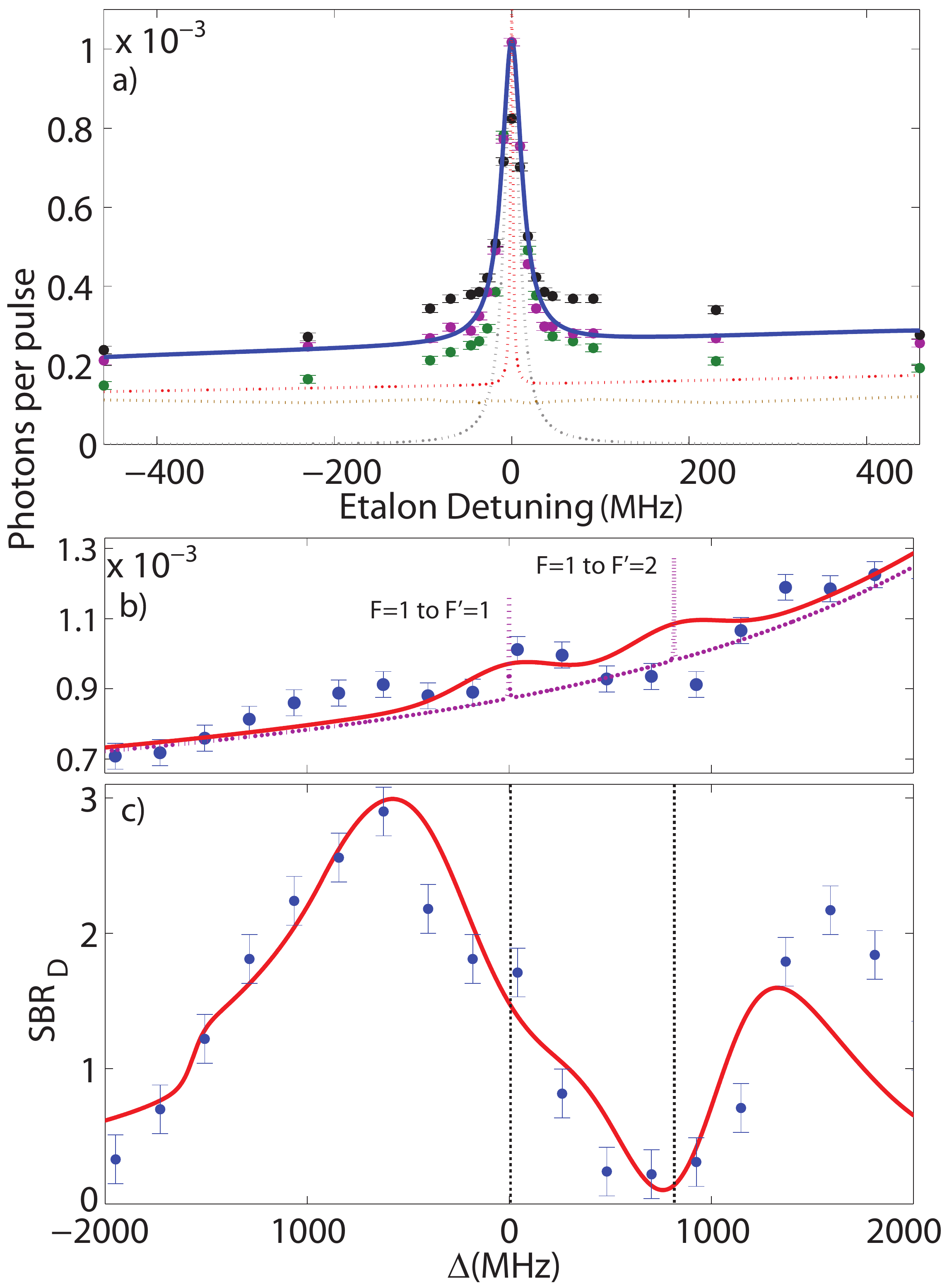}
\caption{(a) Cold atom background response $Q(\Delta)$ (dashed red-line) featuring the contributions of incoherent scattering and Stokes fields; etalon transmission profile (dashed blue line); convoluted response indicating the background transmission through the filtering elements (solid blue line); experimental background measurement for $\Delta= -500$~MHz (green dots), 0~MHz (purple dots) and $+500$~MHz (black dots); technical background (brown dotted line). (b) Cold atom background bandwidths $Q_{1}(\Delta)$ and $Q_{2}(\Delta)$ for the two excited states of the rubidium D1 line manifold (the blue dotted line is a master equation prediction of the background bandwidth); warm atom background response $Q_{RT}(\Delta)$ ((the solid red line is the result of the convolution with a velocity distribution)); background measurements vs. $\Delta$ (blue dots). (c) Predicted room temperature signal to background ratio $SBR_{RT}\propto (\eta_{RT}(\Delta))(T_{RT}(\Delta))/(Q_{RT}(\Delta))$ (solid red line); SBR experimental measurements (blue dots). The error bars are statistical.}
\end{figure}
The background response $Q(\Delta)$ is the combination of two quantum fields. Firstly, from transition $|1\rangle$ to $|3\rangle$, which is narrow and associated to photons incoherently scattered from state $|3\rangle$. This is a result of population exchange with the virtual state $|4\rangle$ mediated by decoherence rates between the ground state $|1\rangle$ and $|2\rangle$. Secondly, from the $|2\rangle$ to $|4\rangle$ transition, which is broad and associated to photons scattered from the virtual state $|4\rangle$ (Stokes field) through an off-resonant Raman process (see dotted red line in Fig. 3a) \cite{Phillips2011,Lauk2013}. These two fields differ by 13.6 GHz.\\
We test our model by detecting background photons passing our filtering elements after exciting the atoms only with control field pulses (fixed $\Delta$, varying etalon detunings, dots in Fig. 3a). These measurements are accurately resembled (see solid blue line in Fig. 3a) by convoluting $Q(\Delta)$ with the etalon transmission function $E(\Delta)=\frac{(1-A)^2}{1+R^2-2R \cos(\frac{2\pi\Delta}{FSR})}$ (dashed blue line in Fig. 3a). The total response is the sum of two convolutions calculated separately for each of the response background components (dotted red line in Fig 3a) and normalized to the input number of background photons before the etalon. We have used R= 0.9955, A=$2*10^{-4}$ and a FSR= 13.6GHz.\\
We obtain the room temperature background response $Q_{RT}(\Delta)$ by considering two background responses $Q_{1}(\Delta)$ and $Q_{2}(\Delta)$ (corresponding to two excited transitions of the rubidium D1 manifold, see blue dotted line in Fig. 3b) and convoluting them with the velocity distribution of the moving atoms (see Fig. 3b red line). This model is in agreement with measurements of the background with fixed etalon detunings and varying $\Delta$. Our final model for the room temperature SBR is calculated as $SBR_{RT}=(\eta_{RT}(\Delta))(T_{RT}(\Delta))/(Q_{RT}(\Delta))$ (solid red line in Fig. 3c) and accurately predicts the features of the SBR measurements, with an optimal operational point corresponding to $\Delta = $ 500 MHz from the central $F=1$ to $F'=1$ resonance.\\
\emph{Ultra low noise storage of polarization qubits.} The predicted optimal performance region is probed by using a one-photon detuning $\Delta \sim$ 250 MHz (red detuned), and storing light pulses with an average $\langle n\rangle=1$ photons and $|H\rangle$ polarization using only a single rail of the setup. The result shows a SBR of $\sim$6 for a storage time of 700 ns and a coherence of a few microseconds (See Fig.4a). Universal qubit operation is verified by using the dual-rail setup sending in and retrieving three sets of orthogonal polarizations, where now the background is inevitably twice that of the single rail. Our outcome was an average qubit SBR of 2.9 $\pm 0.04$ with an average efficiency of 5.1\% $\pm$ 0.07 for the six polarization states $|H\rangle, |V\rangle, |D\rangle, |A\rangle, |R\rangle, |L\rangle$ within a region of interest (ROI) of 400~ns (equal to the input pulse width) upon switching the control field (See Fig. 4b).
\begin{figure}[h!]
\includegraphics[width=\linewidth]{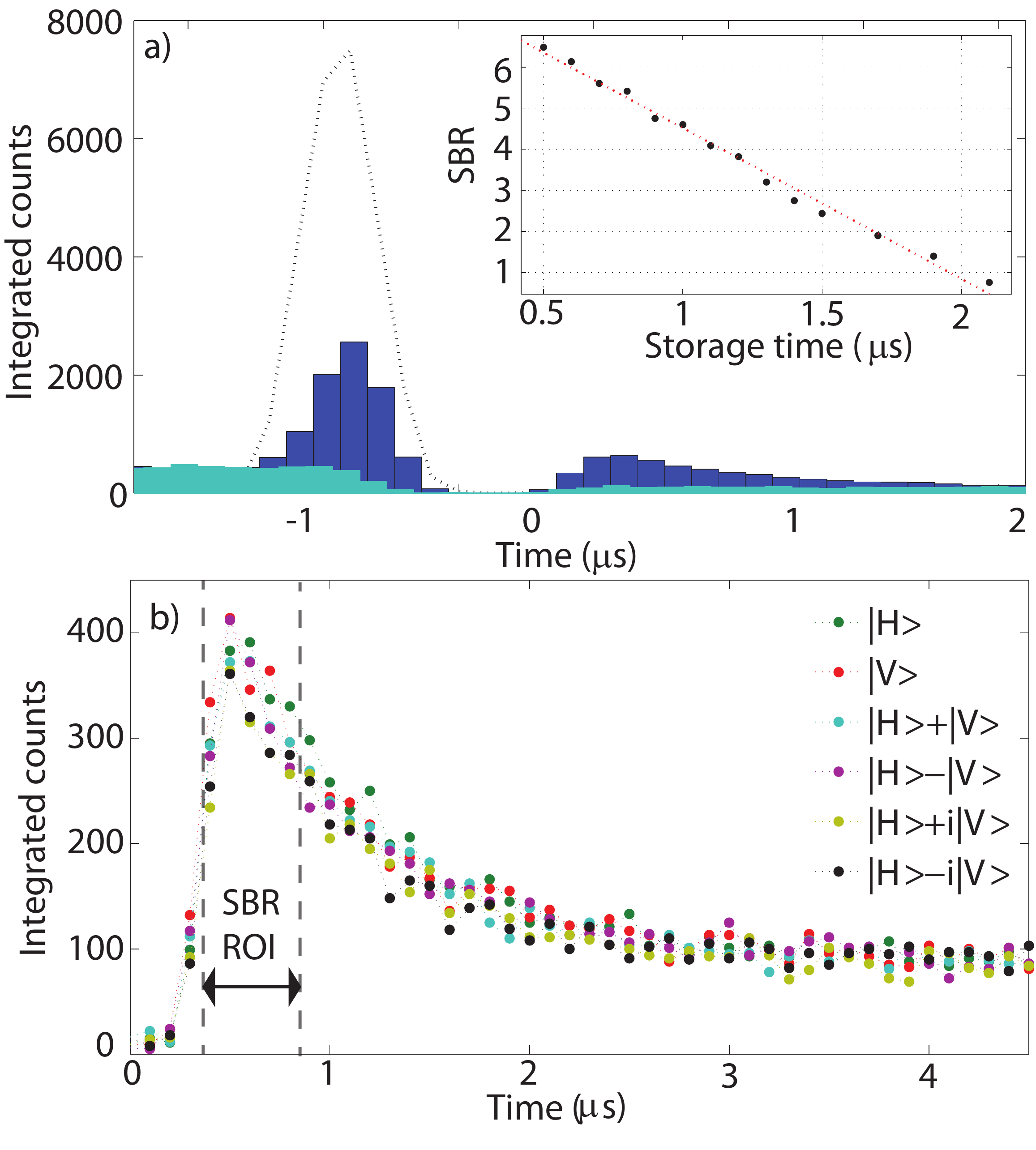}
\caption{(a) Single rail storage with SBR of about 6 where the histogram of photons counts shows the retrieved signal (dark blue bars) compared background counts (light blue bars), the original pulse is shown by the dotted black line. Inset: SBR vs. storage time (black dots) and experimental fit (redline). (b) storage efficiencies for six different input polarizations using the dual rail system.}
\end{figure}
The polarization of each of the retrieved qubit states is obtained with the following procedure \cite{Kupchak2015}: (a) measurement of the polarization of all the input states, (b) qubit storage experiment and determination of the output Stokes vectors ($S_{out}$), (c) rotation of input states to match the orthogonal axis of the normalized stored vectors ($S_{in}$) and (d) evaluation of the total fidelity using $F=\frac{1}{2}(1+\textbf{S}_{out}\cdot \textbf{S}_{in}+\sqrt{(1-\textbf{S}_{out}\cdot \textbf{S}_{out})(1-\textbf{S}_{in}\cdot \textbf{S}_{in})})$. We obtained an average fidelity of $86.6\pm 0.6\%$. This result is well above $71\%$, the fidelity achievable by a classical memory applying the intercept-resent attack and $83.6\%$, the maximum fidelity achievable considering the more elaborate classical strategy exploiting the non-unitary character of the memory efficiency, for a system using attenuated coherent states with $\langle n\rangle=1$ and storage efficiency of $5\%$ \cite{Gundogan2012}.  Furthermore, by reducing the ROI below 400ns, the qubit SBR is improved to 3.7 $\pm 0.09$ corresponding to fidelities of $\sim$ 90\%.\\
\begin{figure}[h!]
\includegraphics[width=\linewidth]{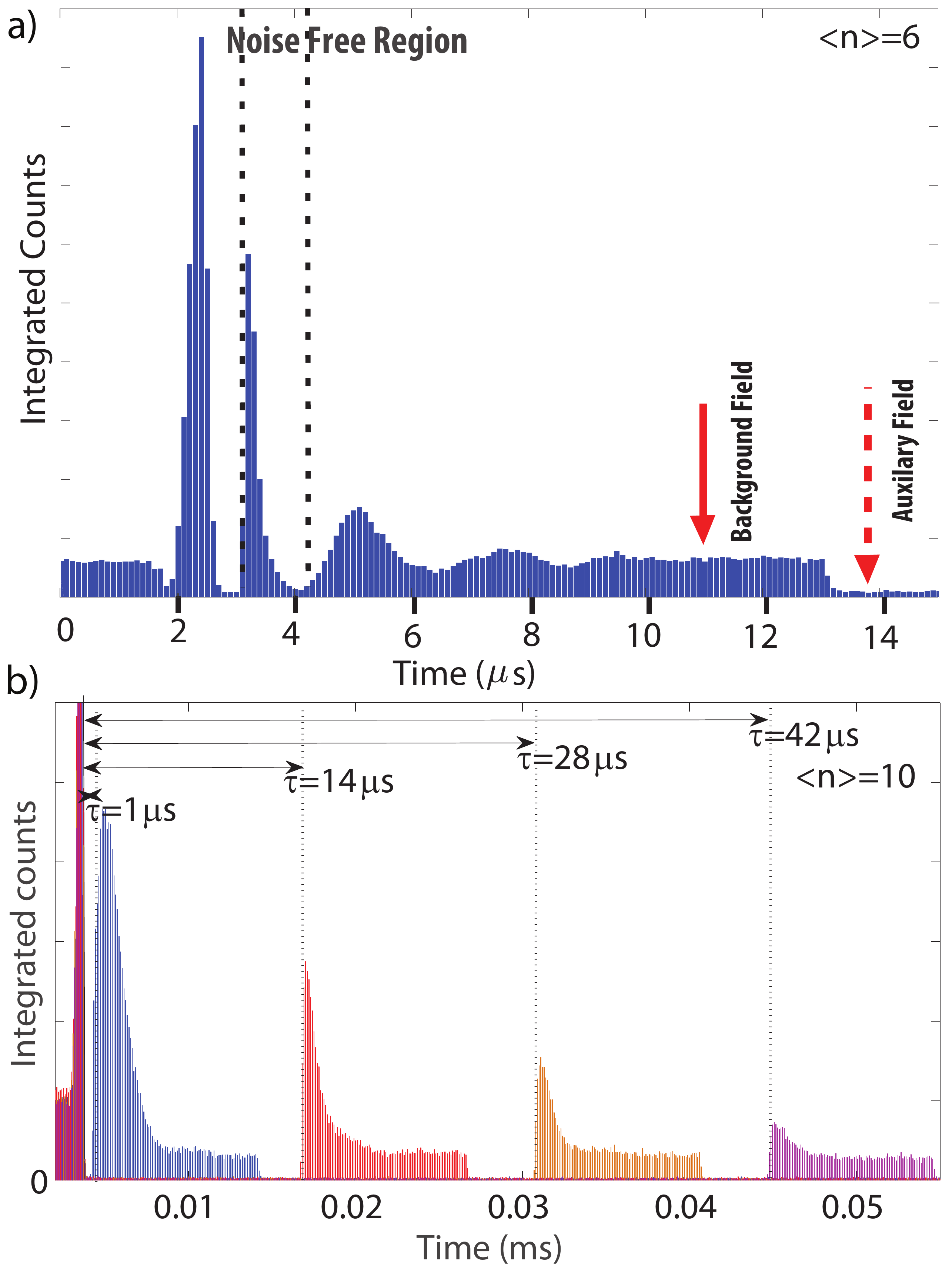}
\caption{a) Noise reduction by introducing an auxiliary field (dashed red), the interaction between dark-state-polaritons creates a background free region. Retrieving the probe within this interval results in a SBR $>$25 for the retrieved probe field. b) Storage of light at the few photon level with storage times $\tau$ =1$\mu$s ($\eta$=11\%, blue),  $\tau$ =14$\mu$s ($\eta$=5.6\%, red), $\tau$ =28$\mu$s ($\eta$=3.1\%, orange) and $\tau$ =42$\mu$s ($\eta$=1.1\%, purple).}
\end{figure}
\noindent \emph{Noise free quantum memory operation.} We have assumed the background response $Q(\Delta)$ to be a combination of two quantum fields produced by different physical mechanism and differing by 13.6 GHz. Moreover, our quantum model also shows that a decrease in the inelastic scattering rate can be engineered by applying simple re-pumping schemes. We have tested these concepts by replacing one of the etalons in the filtering system with a similar unit with different free spectral ratio. This allow us to eliminate the background produced by scattering from the virtual state $|4\rangle$. Furthermore, we have added a weak re-pumping beam on resonance with the $5S_{1/2} F = 1$ $\rightarrow$ $5P_{1/2} F' = 1$ transition that remains on during the complete storage procedure. This results in an oscillatory signal and a complete reduction of the remaining background noise at certain phase relation. We interpret this dynamics as the interaction of two dark-state-polariton modes, one formed by the re-pumper and $\Omega_{c}$, and one by $\Omega_{c}$ and the scattered photons from state $|3\rangle$ \cite{Karpa2008}.\\
Figure 5a shows the obtained oscillations for storage of pulses with $\langle n\rangle = 6$ and an increased control field power in order to highlight the dynamics. By controlling the phases of the fields, it is possible to overlap the retrieved pulse with the noise-free region, translating into a SBR $>25$ and $\eta\sim 15\%$ for the single-photon level case. This predicts a corresponding fidelity of $98\%$, already establishing our system as a viable alternative to cryogenic and cold-atom technologies \cite{Riedl2012,Specht2011,Gundogan2012}. Our results are obtained without the need of either cavity suppression \cite{Saunders2016}, nor ultra-fast pulse operation \cite{Wolters2017,Dou2017,Kaczmarek2017}. Further refinements on these techniques will lead to high-fidelity qubit operation with efficiencies above 50\%.\\
We finish our investigation by improving the achievable storage times. We have used a different cell with a different amount of buffer gas and a low collisional depolarization cross-section (30 Torr Neon) and achieved storage times of $\sim$ 50 $\mu$s at the few-photon-level (see Fig. 5b). By adding anti-relaxation coatings to the interior cell walls, storage times of $\sim 1$~ms are within reach \cite{Seltzer2009}.\\
In conclusion, we have shown the experimental road-map to achieve noise-free room temperature qubit memory operation. Our full quantum analysis of the memory generated background noise made it possible to design and implement techniques to fully suppress it. These important developments allowed us to surpass for the first time with a quantum room temperature device, all-important thresholds related to the performance of the memory in a quantum communication setting.\\
Our realization is already suitable for memory assisted device independent quantum key distribution. This important quantum protocol only needs attenuated coherent states and the relevant parameter is the Quantum Bit Error Rate (QBER, related to the qubit fidelity as $1-F$), which is independent from the memory efficiency \cite{Scarani2009}. In a separate experiment our memory has worked in a shot-by-shot basis when probed with random polarization qubits \cite{Namazi2016}, paving the way for interconnection with polarization entanglement.\\
Together with the development of heralding mechanisms, we envision this technology to become the backbone of future quantum repeater applications based upon outside of the laboratory storage and retrieval of entangled states \cite{Lloyd2001}.\\
E. F. kindly acknowledges A. Neuzner, T. Latka, J. Schupp, S. Ritter and G. Rempe for performing and discussing preliminary experiments on the topic of this paper at the Max Planck Institute of Quantum Optics in Garching, Germany. We also thank I. Novikova for lending us a Rb cell with high buffer as pressure. The work was supported by the US-Navy Office of Naval Research, grant number N00141410801, the National Science Foundation, grant number PHY-1404398 and the Simons Foundation, grant number SBF241180. C. K. acknowledges financial support from the Natural Sciences and Engineering Research Council of Canada. B. J. acknowledges financial assistance of the National Research Foundation (NRF) of South Africa.


\begin{thebibliography}{10}
\expandafter\ifx\csname url\endcsname\relax
  \def\url#1{\texttt{#1}}\fi
\expandafter\ifx\csname urlprefix\endcsname\relax\def\urlprefix{URL }\fi
\providecommand{\bibinfo}[2]{#2}
\providecommand{\eprint}[2][]{\url{#2}}




\bibitem{Ritter2015}
R. Ritter, \emph{et~al.}, Appl. Phys. Lett. \textbf{107}, 041101 (2015).

\bibitem{Borregaard2016}
J. Borregaard, \emph{et~al.}, Nat. Comms. \textbf{7}, 11356 (2016).


\bibitem{DLCZ}
\bibinfo{author}{L.-M. Duan,}, \bibinfo{author}{M.~D. Lukin,},
  \bibinfo{author}{J.~I. Cirac,} \& \bibinfo{author}{P. Zoller,}
\newblock \emph{\bibinfo{journal}{Nature}} \textbf{\bibinfo{volume}{414}},
  \bibinfo{pages}{413} (\bibinfo{year}{2001}).

\bibitem{Abruzzo2014}
S. Abruzzo, H. Kampermann and D. Bruss, Phys. Rev. A \textbf{89}, 012301 (2014).

\bibitem{Panayi2014}
C. Panayi, M. Razavi, X. Ma and N. Luetkenhaus, New J. Phys. \textbf{16}, 043005 (2014).






\bibitem{Lvovsky2009}
\bibinfo{author}{A.~I. Lvovsky,} \bibinfo{author}{B.~C. Sanders,} \&
  \bibinfo{author}{W. Tittel,}
\newblock \emph{\bibinfo{journal}{Nat. Phot.}}
  \textbf{\bibinfo{volume}{3}}, \bibinfo{pages}{706}
  (\bibinfo{year}{2009}).

\bibitem{Bussieres2013}
\bibinfo{author}{F. Bussi{\`e}res,} \emph{et~al.}
\newblock \emph{\bibinfo{journal}{J. Mod. Opt.}}
  \textbf{\bibinfo{volume}{60}}, \bibinfo{pages}{1519}
  (\bibinfo{year}{2013}).

\bibitem{Northup2014}
\bibinfo{author}{T.~E. Northup,} \& \bibinfo{author}{R. Blatt,}
\newblock \emph{\bibinfo{journal}{Nat. Phot.}}
  \textbf{\bibinfo{volume}{8}}, \bibinfo{pages}{356}
  (\bibinfo{year}{2014}).

\bibitem{Reiserer2014}
\bibinfo{author}{A. Reiserer,} \& \bibinfo{author}{G. Rempe,}
\newblock \emph{\bibinfo{journal}{Rev. Mod. Phys.}}
  \textbf{\bibinfo{volume}{87}}, \bibinfo{pages}{1379}
  (\bibinfo{year}{2015}).





\bibitem{Heshami2016}
K. Heshami, \emph{et~al.}, J. Mod. Opt. \textbf{63}, S42 (2016).


\bibitem{Lee2011}
\bibinfo{author}{K.~C. Lee,} \emph{et~al.}
\newblock \emph{\bibinfo{journal}{Science}} \textbf{\bibinfo{volume}{334}},
  \bibinfo{pages}{1253} (\bibinfo{year}{2011}).

\bibitem{Bader2014}
\bibinfo{author}{K. Bader,} \emph{et~al.}
\newblock \emph{\bibinfo{journal}{Nat. Commun.}}
  \textbf{\bibinfo{volume}{5}}, \bibinfo{pages}{5304} (\bibinfo{year}{2014}).

\bibitem{Yao2012}
\bibinfo{author}{N. Yao,} \emph{et~al.}
\newblock \emph{\bibinfo{journal}{Nat. Commun.}}
  \textbf{\bibinfo{volume}{3}}, \bibinfo{pages}{800} (\bibinfo{year}{2012}).



\bibitem{Krauter2013}
\bibinfo{author}{H. Krauter,} \emph{et~al.}
\newblock \emph{\bibinfo{journal}{Nat. Phys.}}
  \textbf{\bibinfo{volume}{9}}, \bibinfo{pages}{400}
  (\bibinfo{year}{2013}).


\bibitem{Jensen2016}
  K. Jensen, \emph{et al}., Sci. Rep. \textbf{6}, 29638 (2016).





















\bibitem{Eisaman2005}
\bibinfo{author}{M.~D. Eisaman,} \emph{et~al.}
\newblock \emph{\bibinfo{journal}{Nature}} \textbf{\bibinfo{volume}{438}},
  \bibinfo{pages}{837} (\bibinfo{year}{2005}).

\bibitem{Reim2011}
\bibinfo{author}{K.~F. Reim,} \emph{et~al.}
\newblock \emph{\bibinfo{journal}{Phys. Rev. Lett.}}
  \textbf{\bibinfo{volume}{107}}  \bibinfo{pages}{053603} (\bibinfo{year}{2011}).




\bibitem{Hosseini2011_2}
\bibinfo{author}{M. Hosseini,}, \bibinfo{author}{G. Campbell,},
  \bibinfo{author}{B.~M. Sparkes,}, \bibinfo{author}{P.~K. Lam,} \&
  \bibinfo{author}{B.~C. Buchler,}
\newblock \emph{\bibinfo{journal}{Nat. Phys.}}
  \textbf{\bibinfo{volume}{7}}, \bibinfo{pages}{794}
  (\bibinfo{year}{2011}).

\bibitem{Sprague2014}
\bibinfo{author}{M.~R. Sprague,} \emph{et~al.}
\newblock \emph{\bibinfo{journal}{Nat. Phot.}}
  \textbf{\bibinfo{volume}{8}}, \bibinfo{pages}{287}
  (\bibinfo{year}{2014}).



\bibitem{Manz2007}
\bibinfo{author}{S. Manz,}, \bibinfo{author}{T. Fernholz,},
  \bibinfo{author}{J. Schmiedmayer,} \& \bibinfo{author}{J.-W. Pan,}
\newblock \emph{\bibinfo{journal}{Phys. Rev. A}}
  \textbf{\bibinfo{volume}{75}}, \bibinfo{pages}{040101(R)} (\bibinfo{year}{2007}).

\bibitem{Phillips2011}
\bibinfo{author}{N.~B. Phillips,}, \bibinfo{author}{A.~V. Gorshkov,} \&
  \bibinfo{author}{I. Novikova,}
\newblock \emph{\bibinfo{journal}{Phys. Rev. A}}
  \textbf{\bibinfo{volume}{83}}, \bibinfo{pages}{063823} (\bibinfo{year}{2011}).


\bibitem{Bashkansky2012}
\bibinfo{author}{M. Bashkansky,}, \bibinfo{author}{F.~K. Fatemi,} \&
  \bibinfo{author}{I. Vurgaftman,}
\newblock \emph{\bibinfo{journal}{Opt. Lett.}}
  \textbf{\bibinfo{volume}{37}}, \bibinfo{pages}{142}
  (\bibinfo{year}{2012}).

\bibitem{Vurgaftman2013}
\bibinfo{author}{I. Vurgaftman,} \& \bibinfo{author}{M. Bashkansky,}
\newblock \emph{\bibinfo{journal}{Phys. Rev. A}}
  \textbf{\bibinfo{volume}{87}}, \bibinfo{pages}{063836} (\bibinfo{year}{2013}).





\bibitem{Kupchak2015}
\bibinfo{author}{C. Kupchak,} \emph{et~al.}
\newblock \emph{\bibinfo{journal}{Sci. Rep.}}
  \textbf{\bibinfo{volume}{5}}, \bibinfo{pages}{7658} (\bibinfo{year}{2015}).

\bibitem{Namazi2015}
\bibinfo{author}{M. Namazi,} \emph{et~al.}
\newblock \emph{\bibinfo{journal}{Phys. Rev. A}}
  \textbf{\bibinfo{volume}{92}}, \bibinfo{pages}{033846} (\bibinfo{year}{2015}).




\bibitem{Lauk2013}
N. Lauk, C. OBrien and M. Fleischhauer, Phys. Rev. A,\textbf{ 88}, 013823 (2013).


\bibitem{Gundogan2012}
\bibinfo{author}{M. Gundogan,}, \bibinfo{author}{P.~M. Ledingham,},
  \bibinfo{author}{A. Almasi,}, \bibinfo{author}{M. Cristiani,} \&
  \bibinfo{author}{H. de~Riedmatten,}
\newblock \emph{\bibinfo{journal}{Phys. Rev. Lett.}}
  \textbf{\bibinfo{volume}{108}}, \bibinfo{pages}{190504} (\bibinfo{year}{2012}).


\bibitem{Karpa2008}
\bibinfo{author}{L. Karpa,} \emph{et~al.}
\newblock \emph{\bibinfo{journal}{Phys. Rev. Lett.}}
  \textbf{\bibinfo{volume}{101}}, \bibinfo{pages}{170406}
  (\bibinfo{year}{2008}).

\bibitem{Saunders2016}
D.~J. Saunders, \emph{et~al.}, Phys. Rev. Lett. \textbf{116}, 090501 (2016).

\bibitem{Wolters2017}
J. Wolters,  \emph{et~al.}, arXiv:1703.00489 (2017).

\bibitem{Dou2017}
J. P. Dou,  \emph{et~al.}, arXiv:1704.06309 (2017).

\bibitem{Kaczmarek2017}
K. T. Kaczmarek,  \emph{et~al.}, arXiv:1704.00013 (2017).


\bibitem{Scarani2009}
\bibinfo{author}{V. Scarani,} \emph{et~al.}
\newblock \emph{\bibinfo{journal}{Rev. Mod. Phys.}}
  \textbf{\bibinfo{volume}{81}}, \bibinfo{pages}{1301}
  (\bibinfo{year}{2009}).


\bibitem{Riedl2012}
\bibinfo{author}{S. Riedl,} \emph{et~al.}
\newblock \emph{\bibinfo{journal}{Physical Review A}}
  \textbf{\bibinfo{volume}{85}}, \bibinfo{pages}{022318} (\bibinfo{year}{2012}).

  \bibitem{Specht2011}
\bibinfo{author}{H.~P. Specht,} \emph{et~al.}
\newblock \emph{\bibinfo{journal}{Nature}} \textbf{\bibinfo{volume}{473}},
  \bibinfo{pages}{190} (\bibinfo{year}{2011}).

\bibitem{Namazi2016}
Mehdi Namazi, \emph{et~al.}, arXiv:1609.08676 (2016).

\bibitem{Seltzer2009}
S. J. Seltzer and M. V. Romalis, J. Appl. Phys. \textbf{106}, 114905 (2009).

\bibitem{Lloyd2001}
S. Lloyd, \emph{et~al.}, Phys. Rev. Lett. 87, 167903 (2001).

\end{thebibliography}
\end{document}